# Single crystals of LnFeAsO$_{1-x}$F$_x$ (Ln=La, Pr, Nd, Sm, Gd) and Ba$_{1-x}$Rb$_x$Fe$_2$As$_2$: growth, structure and superconducting properties

J. Karpinski,[1] N. D. Zhigadlo,[1] S. Katrych,[1] Z. Bukowski,[1] P. Moll,[1] S. Weyeneth,[2] H. Keller,[2] R. Puzniak,[3] M. Tortello,[4] D. Daghero,[4] R. Gonnelli,[4] I. Maggio-Aprile,[5] Y. Fasano,[5,6] Ø. Fischer,[5] B. Batlogg[1]

[1]Laboratory for Solid State Physics, ETH Zurich, 8093 Zurich, Switzerland
[2]Physik-Institut der Universität Zürich, 8057 Zürich, Switzerland
[3]Institute of Physics, Polish Academy of Sciences, Aleja Lotników 32/46, 02-668 Warsaw, Poland
[4]Dipartimento di Fisica, Politecnico di Torino, 10129 Torino, Italy
[5]DPMC-MaNEP, University of Geneva, Geneva, Switzerland
[6]Low Temperatures Laboratory and Instituto Balseiro, Bariloche, Argentina

**Abstract**

A review of our investigations on single crystals of LnFeAsO$_{1-x}$F$_x$ (Ln=La, Pr, Nd, Sm, Gd) and Ba$_{1-x}$Rb$_x$Fe$_2$As$_2$ is presented. A high pressure technique has been applied for the growth of LnFeAsO$_{1-x}$F$_x$ crystals, while Ba$_{1-x}$Rb$_x$Fe$_2$As$_2$ crystals were grown using quartz ampoule method. Single crystals were used for electrical transport, structure, magnetic torque and spectroscopic studies. Investigations of the crystal structure confirmed high structural perfection and show less than full occupation of the (O, F) position in superconducting LnFeAsO$_{1-x}$F$_x$ crystals. Resistivity measurements on LnFeAsO$_{1-x}$F$_x$ crystals show a significant broadening of the transition in high magnetic fields, whereas the resistive transition in Ba$_{1-x}$Rb$_x$Fe$_2$As$_2$ simply shifts to lower temperature. Critical current density for both compounds is relatively high and exceeds $2\times10^9$ A/m$^2$ at 15 K in 7 T. The anisotropy of magnetic penetration depth, measured on LnFeAsO$_{1-x}$F$_x$ crystals by torque magnetometry is temperature dependent and apparently larger than the anisotropy of the upper critical field. Ba$_{1-x}$Rb$_x$Fe$_2$As$_2$ crystals are electronically significantly less anisotropic. Point-Contact Andreev-Reflection spectroscopy indicates the existence of two energy gaps in LnFeAsO$_{1-x}$F$_x$. Scanning Tunneling Spectroscopy reveals in addition to a superconducting gap, also some feature at high energy (~20 meV).

# I. Introduction

Since the first report on superconductivity at 26 K in F-doped LaFeAsO at the end of February 2008, the superconducting transition temperature has been quickly raised to about 55 K and several new superconductors of a general formula $LnFeAsO_{1-x}F_x$ (Ln=La, Ce, Pr, Nd, Sm, Gd, Tb, Dy), abbreviated as Ln1111, have been synthesized [1-9]. These compounds crystallize with the tetragonal layered ZrCuSiAs structure, in the space group of $P4/nmm$. The structure consists of alternating LnO and FeAs layers, which are electrically charged represented as $(LnO)^{+\delta}(FeAs)^{-\delta}$ (Fig. 1). Covalent bonding is dominant in the layers, while ionic bonding dominates between layers. Electron carriers can be introduced by substituting F for O or by oxygen deficiency [1-8]. By substituting $Sr^{2+}$ for $La^{3+}$ in La1111, holes are introduced.

More recently, superconductivity in $AFe_2As_2$ (A=Ca, Sr, Ba) (called A122) with $ThCr_2Si_2$-type structure and maximum $T_c$ = 38 K has been reported [10]. These compounds have a more simple crystal structure in which $(Fe_2As_2)$-layers, identical to those in Ln1111 are separated by single elemental A layers (Fig. 2). Up to date superconductivity has been found in hole-doped $Sr_{1-x}K_xFe_2As_2$ and $Sr_{1-x}Cs_xFe_2As_2$ [11], $Ca_{1-x}Na_xFe_2As_2$ [12], $Eu_{1-x}K_xFe_2As_2$ [13], and $Eu_{1-x}Na_xFe_2As_2$ [14], as well as in electron-doped Co-substituted $BaFe_2As_2$ [15] and $SrFe_2As_2$ [16], and Ni-substituted $BaFe_2As_2$ [17]. Furthermore, pressure induced superconductivity has been also discovered in the parent compounds $CaFe_2As_2$ [18, 19], $SrFe_2As_2$ [20, 21], and $BaFe_2As_2$ [21].

Besides $KFe_2As_2$ and $CsFe_2As_2$, which are superconductors with $T_c$'s of 3.8 K and 2.6 K [11] respectively, $RbFe_2As_2$ is known to exist as well [22]. Therefore, it seemed natural to us to explore the $BaFe_2As_2$-$RbFe_2As_2$ system in order to search for superconductivity.

It is interesting to explore the important parameters which govern the superconducting properties of new superconductors. In the case of high-$T_c$ cuprates it is the number of carriers doped into the $CuO_2$ layers. In analog in the pnictides it is the number of carriers doped into the FeAs layers. There are some similarities between the new pnictide superconductors and the cuprate superconductors due to the layered structure and the fact that both Fe and Cu are $3d$ elements. However, there are important differences. First, doping on the Fe site in Sr122 or Ba122 by substitution of Fe by Co leads also to the appearance of superconductivity [15, 16] in contrast to cuprates, where substitution for Cu suppresses superconductivity. Second, in cuprates, introducing of one oxygen atom is equivalent to introducing of two fluorine atoms [23]. In Ln1111 there is a significant difference in carrier doping between oxygen deficiency and fluorine substitution. One expects that one oxygen atom deficiency provides two electrons while substitution of $F^-$ for $O^{2-}$ provides one electron. However, according to [24, 25] oxygen deficiency is much less effective as a source of electrons than F-substitution. Structural parameters play also an important role for obtaining high $T_c$'s. There is dependence between $T_c$ and the As-Fe-As bond angle of the $FeAs_4$ tetrahedron: maximum $T_c$ is achieved when As-Fe-As bond angle is close to 109.47° corresponding to an ideal tetrahedron [25].

Despite all these differences, the similarities due to the layered crystal structure are important as well. So far, all high-$T_c$ superconductors have a layered crystal structure leading to pronounced anisotropic physical properties. All cuprate superconductors have been characterized by a well-defined effective mass anisotropy parameter γ [26]:

$$\gamma = \sqrt{m_c^*/m_{ab}^*} = \lambda_c/\lambda_{ab} = \xi_{ab}/\xi_c = H_{c2}^{||ab}/H_{c2}^{||c}. \quad (1)$$

where $m^*_i$ denote the effective mass, $\lambda_i$ the magnetic penetration depth, $\xi_i$ the coherence length and $H_{c2}^{||i}$ the upper critical field in the magnetic field direction $i$. Nevertheless, the understanding of high temperature superconductivity was challenged by the observation of two distinctly different and temperature dependent anisotropies in $MgB_2$ single crystals [27, 28, 29]:

$$\gamma_\lambda = \lambda_c/\lambda_{ab} \quad (2)$$

$$\gamma_H = H_{c2}^{||ab}/H_{c2}^{||c} \quad (3)$$

A straight forward interpretation based on a two-band model was quickly developed, which also lead to a further understanding of the temperature and field dependence of the anisotropy parameters in $MgB_2$, mirroring the complex inter- and intraband mechanism of the two superconducting gaps [30]. For an overall comparison with the other high-temperature superconductors (e. g. cuprates, $MgB_2$) a detailed knowledge of γ in the oxypnictides is required. Various attempts were made to determine the actual anisotropy in the oxypnictide superconductors [31-42], leading to a wide range of results. Nevertheless, from both experimental and theoretical sides there is clear evidence that superconductivity in the pnictides involves more than one band [31-33, 43-48].

Not only the anisotropic properties are, obviously, best investigated on single crystals. Single crystals are also required for spectroscopic techniques such as Scanning Tunneling

Spectroscopy (STS), Angle-Resolved Photoemission Spectroscopy (ARPES), Point-Contact Andreev-Reflection (PCAR) spectroscopy, optical spectroscopy, etc. A number of recent investigations of oxypnictides have focused on the multiband superconductivity [31-33, 43-48]. Answering the question of whether the different Fermi surface sheets are associated with different gaps is of crucial importance in order to identify the mechanisms of superconductivity in these compounds. In this regard, experimental techniques such as ARPES, PCAR and STS are very powerful methods since they allow a direct determination of the energy gap(s). STS is a suitable technique to study this issue since it probes the quasiparticle excitation spectrum in the superconducting state, a direct measure of the local density of states and therefore of the fundamental properties of the superconducting order parameter. This technique was successfully applied to $MgB_2$, where multiband superconductivity was unambiguously demonstrated through directional STS measurements [49].

We succeeded in growing of the first free standing FeAs-oxypnictides crystals ($SmFeAsO_{1-x}F_y$) using a high-pressure technique and NaCl/KCl flux [50]. The NaCl/KCl flux has very low solubility at temperatures below 1000 °C used for processes in quartz ampoules, therefore crystal growth at this temperature is extremely slow [51]. In order to increase the solubility in NaCl/KCl flux for more efficient crystal growth higher temperature should be used, but Ln1111 becomes unstable. This trend can be counteracted by applying high pressure, which then tends to stabilize the structure of Ln1111 at high temperature.

Single crystals of $AFe_2As_2$ can be grown from Sn flux, similar to many other intermetallic compounds [52, 53]. Tin is practically the only metal that dissolves iron reasonably well and does not form stable unwanted compounds. Due to high solubility in Sn flux at temperatures compatible with quartz ampoules, large, millimeter-sized crystals of A122 have been grown, which allowed extensive measurements of their physical properties. The disadvantage of the Sn-flux technique is that crystals usually contain ~1% at. Sn. Another method of growing $AFe_2As_2$ crystals is the high-temperature growth from FeAs flux [54].

Here, we report on the crystal growth using both the high-pressure, high-temperature method with NaCl/KCl flux for Ln1111 and the quartz ampoule method with Sn flux for A122 [55]. The results of structure investigations on series of Ln1111 crystals (Ln=Sm, Nd, Pr, La, Gd) and $Ba_{1-x}Rb_xFe_2As_2$ are presented. Electrical resistivity measurements, investigations of the anisotropy parameter and spectroscopic studies are also summarized.

**II. Experimental**

## 1. Crystal growth

For the synthesis of LnFeAsO$_{1-x}$F$_x$ (Ln=La, Pr, Nd, Sm, Gd) polycrystalline samples and single crystals we used a cubic anvil high-pressure technique which has been successfully applied in our laboratory at ETH Zurich also for the single crystal growth of MgB$_2$ and other superconductors. The mixture of LnAs, FeAs, Fe$_2$O$_3$, Fe, and LnF$_3$ powders was used as a precursor. For the growth of single crystals we used additionally NaCl/KCl flux. The precursor-to-flux ratio varies between 1:1 and 1:3. By variation of nominal oxygen and fluorine content between 0.6-0.8 and 0.4-0.2 respectively different doping levels were achieved. The precursor powders were ground mixed and pressed into pellets in a glove box due to toxicity of arsenic. Pellets containing precursor and flux were placed in a BN crucible inside a pyrophyllite cube with a graphite heater. Six tungsten carbide anvils generated pressure on the whole assembly. In a typical run, a pressure of 3 GPa was applied at room temperature. For the crystal growth the temperature was increased within 1 h to the maximum value of 1350-1450 °C, kept for 4-85 h and decreased in 1-24 h to room temperature. For the synthesis of polycrystalline samples the maximum temperature of 1300-1350 °C was kept for 2-6 h followed by quenching. Then the pressure was released, the sample removed and in the case of single crystal growth the NaCl/KCl flux dissolved in water. After drying, the shiny single crystals could be selected easily. One has to mention that such high-pressure experiments have to be performed very carefully, because an explosion during heating due to increased pressure in the sample container can lead to a contamination of the whole apparatus with arsenide compounds.

With the aim of growing single crystals suitable for physical measurements, we carried out systematic investigations of the parameters controlling the growth of crystals, including growth temperature, applied pressure, starting composition, dwelling time and heating/cooling rate. Most of our exploratory crystal growth experiments were performed on the SmFeAs(O, F) system. Figure 3 shows typical single crystals of SmFeAs(O, F), obtained in the growth experiment at 30 kbar and 1380 °C for 60-85 h. Their size is much smaller than the size of single crystals of Ba$_{1-x}$Rb$_x$Fe$_2$As$_2$ (Fig. 4). By optimization of the growth conditions SmFeAs(O, F) single crystals with the sizes in the range of 150-300 μm and $T_c \approx 53$ K have been obtained. In general, extending the soak time leads to larger crystals, but parasitic phases such as FeAs balls are formed simultaneously. One of the problems of crystal growth at high-temperature and high-pressure conditions is that the density of sites for nucleation is high and it is difficult to control nucleation so that fewer, but larger crystals would grow. This is also reflected on the quality of the grown crystals, therefore many of them have irregular shapes and they form clusters of several crystals. The solubility of SmFeAs(O, F) in NaCl/KCl flux is very low, which results in small crystals. In order to obtain larger and high quality crystals, it is important to choose the growth condition where crystals grow from only limited number of nuclei at a reasonable growth rate. It is necessary to search for a new solvent system with higher solubility and in which the kinetic barrier for nucleation of the Ln1111-phase is large. The existence of parasitic phases also has a significant effect on the growth mechanism and appropriate doping. Too high precursor to flux ratio prevents growth of larger crystals because of insufficient space for growth of individual grains. The crystal growth conditions which were optimized for the growth of optimally doped and relatively large SmFeAs(O, F) crystals have been applied to other systems, such as Nd-, La-, Gd-, and PrFeAs(O, F). For all these system we were able to grow single crystals. For each of these compounds, however the growth conditions and composition of the precursor need to be optimized. X-ray diffraction analysis of high pressure crystal growth products indicate that several parasitic reactions proceeded in the crucible together with the single-crystal growth of the Ln1111-type phase. For example, in the case of the LaFeAs(O, F) system, we observed several impurity phases, such as FeAs, LaOF, LaOCl, etc. $T_c$'s measured with a SQUID magnetometer on Ln1111 (Ln = La, Pr, Nd, Sm, Gd) single crystals with various F doping are presented in Fig. 5.

Single crystals of $Ba_{1-x}Rb_xFe_2As_2$ were grown using a Sn flux method similar to that described in Ref. [54]. The Fe:Sn ratio (1:24) in a starting composition was kept constant in all runs while the Rb:Ba ratio was varied between 0.7 and 2.0. The appropriate amounts of Ba, Rb, $Fe_2As$, As, and Sn were placed in alumina crucibles and sealed in silica tubes under 1/3 atmosphere of Ar gas. Next, the ampoules were kept at 850 °C for 3 hours until all components were completely melted, and cooled over 50 hours to 500 °C. At this temperature the liquid Sn was decanted from the crystals. The remaining thin film of Sn at the crystal surfaces was subsequently dissolved at room temperature using liquid Hg, and finally the crystals were heated to 190 °C in vacuum to evaporate the remaining traces of Hg. No signs of superconducting Hg are seen in the magnetic measurements.

The single crystals of $Ba_{1-x}Rb_xFe_2As_2$ grow in a plate-like shape with typical dimensions (1–3) x (1–2) x (0.05–0.1) mm$^3$ (Fig. 4). Depending on the starting composition, the crystals displayed a broad variety of properties from nonsuperconducting to superconducting with sharp transitions to the superconducting state. For further studies we chose single crystals grown from initial composition $Ba_{0.6}Rb_{0.8}Fe_2As_2$. The composition of the crystals from this batch determined by EDX analysis (16.79 at. % Ba, 1.94 at. % Rb, 1.74 at. % Sn, 40.19 at. % Fe, and 39.33 at. % As) leads to the chemical formula $Ba_{0.84}Rb_{0.10}Sn_{0.09}Fe_2As_{1.96}$. Crystals from the selected batch exhibit a moderate $T_c$ (around 22-24 K) but compared to the crystals with higher $T_c$ their superconducting transition is relatively sharp, with more than one step in the resistance curve, however.

## 2. Experimental details of structure and superconducting properties studies

Single crystals were studied on a four-circle diffractometer equipped with CCD detector (X-calibur PX, Oxford Diffraction) using Mo $K_\alpha$ radiation. The single crystals of various batches have been characterized by X-ray diffraction showing well-resolved reflection patterns indicating a high quality of the crystallographic structure. Data reduction and analytical absorption correction were done using the program CrysAlis [56]. The crystal structure was determined by a direct method and refined on $F^2$ employing the programs SHELXS-97 and SHELXL-97 [57, 58].

Magnetic measurements were performed using a Quantum Design SQUID Magnetometer MPMS XL with a standard Reciprocating Sample Option installed. Low field susceptibility measurements revealed a narrow and well-defined transition from the normal to the superconducting state.

Torque magnetometry has been applied to determine γ, a technique which allows to measure the angular dependent superconducting magnetization by detecting the torque of a single crystal in a magnetic field along a certain orientation with respect to the crystallographic *c*-axis. A home made piezoresistive torque sensor was used [59]. The crystal was mounted in an Oxford flow cryostat allowing stabilization of temperatures between 10 K and 300 K. A turnable Bruker NMR magnet with a maximum field of 1.4 T was used to vary the field magnitude and its orientation with respect to the crystallographic axes allowing for a full rotation through 360 degrees. For this experiment several single crystals have been chosen with the nominal composition $SmFeAsO_{0.8}F_{0.2}$ and $NdFeAsO_{0.8}F_{0.2}$ with masses of ~100 ng and $T_c$ of the order of 45 K [51].

Direct four-point resistivity measurements were performed on SmFeAs(O, F) (Sm1111) and (Ba, Rb)$Fe_2As_2$ ((Ba, Rb)122) crystals using a Quantum Design Physical Property Measurement System (PPMS) in magnetic fields up to 14 T. To minimize the broadening of the transition due to material inhomogeneities, Sm1111 crystals smaller than 200 μm were selected and contacted using a Focused Ion Beam. This technique produces precisely deposited micrometer-sized Pt leads onto the crystal and was found not to alter the bulk superconducting properties. The diameter of typical (Ba, Rb)122 crystals was well above 1 mm and allowed for standard manual contacting.

PCAR spectroscopy measurements were performed in SmFeAs(O, F) crystals in order to study the superconducting energy gap(s). The relatively small size of the crystals prevents the use of the standard point-contact technique which consists in pressing a sharp metallic tip against the material under study and also of the "soft" PCAR technique [60] where the contact is achieved by means of a small drop of Ag conductive paste. Instead the current was therefore injected in the crystals through a very small (10 μm diameter) Au-wire which acts as the tip. The crystals were vertically mounted by contacting the lateral edges by means of In. The Au-wire was leant transversely on the thin lateral side of the crystal. In this way, the current is mainly injected along the *ab*-plane (*ab*-plane contact) and it is possible to measure the differential conductance curves, d$I$/d$V$ vs. $V$, across the Au/crystal junction.

STS measurements have been performed on a single crystal of SmFeAsO$_{0.86-x}$F$_x$. Due to the small sizes of the samples (~50x50 μm$^2$), in-situ cleaving or fracturing was not feasible, and therefore all measurements were made on as-grown surfaces. Before being introduced into the UHV chamber, a batch of single crystals was glued on the sample holder, and rinsed in pure deionized water and isopropyl alcohol. Tunneling topographic and spectroscopic measurements were performed with a home-made low-temperature scanning tunneling microscope, at 4.2 K in 10$^{-3}$ mbar He exchange gas pressure. Electrochemically etched Iridium tips served as the ground electrode and were positioned perpendicularly to the (001) face of the crystal. The tunneling resistance was adjusted in the 500 MΩ range (0.1V sample bias voltage, and 200 pA tunnel current) to ensure a true tunneling regime.

### III. Results and discussion

### 1. Crystal structure
### i. Crystal structure of LnFeAs(O, F)

All atomic positions were found using the direct method. The refinement was performed without any constraints. The oxygen and fluorine atoms occupy the same position and were treated as one atom because it is impossible to distinguish between them by X-ray diffraction. The results of the structure refinement are presented in the Table 1. The lanthanide contraction reflects itself in a systematic lattice parameters reduction across the series (Fig. 6 and 7). The only two variable atomic coordinates $z$ for As and Ln also vary regularly across lanthanides series (Fig. 8). All samples reveal more than 10 % vacancies on the O(F) site (Tab. 1). However, the accuracy of the determination of the oxygen/fluorine occupancy is low due to the presence of the heavy As, Fe and Sm elements. The bonding in the layers Ln-O and Fe-As for the LnFeAs(O, F) is covalent. The absolute values of these distances ($d_{Ln-O}$ ~ 2.28-2.34 Å) are close to the sum of covalent radii of the elements ($r^{cov}_{Ln}$ + $r^{cov}_{O}$ ~ 2.27-2.35 Å for Gd-La series). The sum of the covalent radii of the Ln and As atoms ($r^{cov}_{Ln}$ + $r^{cov}_{As}$ ~ 2.82-2.90 Å for Gd-La series) is smaller than the distances between these atoms ($d_{Ln-As}$ ~ 3.24-3.34 Å (Tab. 1). Between the layers ionic bonding is dominant.

**ii. Crystal structure of (Ba, Rb)Fe$_2$As$_2$**

A structure analysis was performed on Rb-substituted BaFe$_2$As$_2$. We assumed that Rb substitute for Ba atoms, and the Rb/Ba occupation was refined simultaneously. Anisotropic displacement parameters as well as atomic coordinates for both elements were restrained to the same value. After several cycles of refinement we found in the Fourier difference map a pronounced maximum close to but displaced from the Ba/Rb site. According to EDX analysis a small amounts of Sn (from the flux) are present in the crystal. We assume that the maximum off the Ba/Rb site corresponds to the location of Sn in the structure. The next step of the refinement was performed with Sn located in the position of maxima, and the occupation parameter of Sn was refined. Inserting Sn in the refinement decreased the $R$ factor considerably (from 5.41 % to 3.89 %). We assumed the overall occupation of Ba, Rb and Sn to be 100 %. The overall occupation of the Rb/Ba site was decreased by the amount of Sn and fixed while the ratio Ba/Rb was refined. After several refinement cycles the absorption correction for the correct crystallographic composition was performed. The occupation parameters for the Rb/Ba and Sn sites were found to be 0.89/0.05 and 0.06, respectively. Therefore, the more appropriate chemical formula is Ba$_{0.89}$Rb$_{0.05}$Sn$_{0.06}$As$_2$Fe$_2$. The results of the final structure refinement are presented in Tab. 2 and the resulting structure in Fig. 9. Compared to unsubstituted BaAs$_2$Fe$_2$ the lattice parameter $a$ is slightly shorter, the $c$ parameter is longer and the volume of the unit cell is smaller. A similar tendency has been observed for other A122 compounds, when Ba or Sr is replaced by K. The increase of the $c$ parameter in $Ba_{0.89}Rb_{0.05}Sn_{0.06}As_2Fe_2$ is caused mainly by substitution of $Ba^{2+}$ ions ($r$ = 1.42 Å) by larger $Rb^+$ ($r$ = 1.61 Å). The relatively large contraction of the *a* parameter (larger than expected from Vegard's law) seems to be the effect of Sn incorporation.

Tab. 1. Details of the structure refinement for the LnFeAs(O, F) (Ln=La, Pr, Nd, Sm, Gd) crystals. The diffraction study is performed at 295(2) K using Mo K$_\alpha$ radiation with λ = 0.71073 Å, The lattice is tetragonal, *P*4/nmm space group with Z=2. The absorption correction was done analytically. A full-matrix least-squares method was employed to optimize $F^2$

| Empirical formula | LaFeAsO$_{0.88-x}$F$_x$ Δ$x$=±2.4 | PrFeAsO$_{0.80-x}$F$_x$ Δ$x$=±2 | NdFeAsO$_{0.89-x}$F$_x$ Δ$x$=±2 | SmFeAsO$_{0.86-x}$F$_x$ Δ$x$=±4 | GdFeAsO$_{0.76-x}$F$_x$ Δ$x$=±4 |
|---|---|---|---|---|---|
| $T_c/\Delta T_c$, K | | 38.8/3 | 46.3/3 | 47.9/2.5 | 22.7/5 |
| Unit cell dimensions (Å) | $a$= 4.02690(10), $c$= 8.7010(3) | $a$= 3.97820(10), $c$= 8.5810(4) | $a$= 3.95940(10), $c$= 8.5443(2) | $a$= 3.93110(10), $c$= 8.4655(5) | $a$= 3.91610(10), $c$= 8.4486(6) |
| Volume (Å$^3$) | 141.095(7) | 135.804(8) | 133.948(6) | 130.822(9) | 129.566(10) |
| Calculated density (g/cm$^3$) | 6.741 | 7.016 | 7.238 | 7.551 | 7.753 |
| Absorption coefficient (mm$^{-1}$) | 31.405 | 34.831 | 36.516 | 39.985 | 43.3 |
| Crystal size (μm$^3$) | 163 x 115 x 24 | 137 x 111 x 12 | 91 x 72 x 10 | 86 x 96 x 14 | 117 x 77 x 18 |
| Θ range for data collection | 4.69 to 42.32 deg | 4.75 to 45.78 deg | 4.77 to 53.87 deg | 4.82 to 40.78 deg | 5.74 to 40.65 deg |
| Index ranges | -7=h<=3, -2<=k<=7, -16<=l<=10 | -6=h<=7, -7<=k<=7, -17<=l<=16 | -8=h<=7, -5<=k<=8, -19<=l<=10 | -5=h<=6, -6<=k<=7, -13<=l<=15 | -6=h<=7, -5<=k<=7, -15<=l<=14 |
| Reflections collected/unique | 1212 /330 $R_{int.}$= 0.0293 | 1612/384 $R_{int.}$= 0.0242 | 1859/474 $R_{int.}$= 0.0264 | 1163/285 $R_{int.}$= 0.0370 | 975/281 $R_{int.}$= 0.0396 |
| Data/restraints/parameters | 330/0/12 | 384/0/12 | 474/0/12 | 285/0/12 | 281/0/12 |
| Goodness-of-fit on $F^2$ | 1.118 | 1.173 | 0.996 | 1.149 | 1.104 |
| Final R indices [$I$>2Ω($I$)] | $R_1$ = 0.0408, w$R_2$ = 0.1145 | $R_1$ = 0.0323, w$R_2$ = 0.0786 | $R_1$ = 0.0286, w$R_2$ = 0.0681 | $R_1$ = 0.0424, w$R_2$ = 0.1172 | $R_1$ = 0.0497, w$R_2$ = 0.1157 |
| $R$ indices (all data) | $R_1$ = 0.0494, w$R_2$ = 0.1190 | $R_1$ = 0.0385, w$R_2$ = 0.0800 | $R_1$ = 0.0393, w$R_2$ = 0.0708 | $R_1$ = 0.0479, w$R_2$ = 0.1199 | $R_1$ = 0.0593, w$R_2$ = 0.1195 |
| Fractional atomic coordinates, O(F) occupation and atomic displacement parameters (Å$^2$) | | | | | |
| Fe $x$ = ¼; $y$ = ¾ ; $z$= ½ ; O(F) ¼; ¾; 0 | | | | | |
| Ln $x$ = -¼; $y$ = -¼  $z$ | 0.1468(1) | 0.1435(1) | 0.1440(1) | 0.1419(1) | 0.1382(1) |
| As $x$ = ¼, $y$ = ¼   $z$ | 0.3474(1) | 0.3435(1) | 0.3414(1) | 0.3388(2) | 0.3375(2) |
| O(F) occupation | 0.88 | 0.80 | 0.89 | 0.86 | 0.76 |
| Interatomic distances (Å) | | | | | |
| Ln-O | 2.3844(5) | 2.3392(2) | 2.3309(1) | 2.3035(4) | 2.2797(5) |
| Ln-As | 3.3398(6) | 3.2952(5) | 3.2686(4) | 3.2411(1) | 3.241(1) |
| Fe-As | 2.4118(5) | 2.4002(5) | 2.3989(4) | 2.393(1) | 2.391(2) |

Table 2. Crystal data for the $Ba_{0.89}Rb_{0.05}Sn_{0.06}Fe_2As_2$.

| | |
|---|---|
| Crystal system, space group, Z | Tetragonal, *I*4/mmm, 2 |
| Unit cell dimensions (Å) | *a*= 3.9250(2), *c*= 13.2096(5) |
| Volume (Å$^3$) | 203.502(3) |
| Fractional atomic coordinates | |
| Ba/Rb: *x*=*y*=*z*=0, Sn: *x*=*y*=0; *z*=0.0837(7), As: *x*=*y*=0; *z*=0.3543(1), | |
| Fe: *x*=1/2; *y*=0; *z*=1/4 | |
| Bond lengths (Å) | |
| Ba/Rb-As | 3.3774(3) x 8 |
| Fe-As | 2.3979(3) x 4 |
| Fe-Fe | 2.7754(1) x 4 |
| As-Sn | 2.894(3) x 4 |
| Fe-Sn | 2.945(7) x 4 |
| Bond angles (deg) | |
| As-Fe-As | 109.86(2) |
| | 109.28(1) |

## 2. Upper critical fields, critical current and superconducting state anisotropy
### i. Resistivity measurements

Resistivity measurements ρ(*T*, *H*) near $T_c$ for magnetic fields parallel (*H*||*ab*) and perpendicular (*H*||*c*) to the FeAs-planes show remarkably different behavior for Sm1111 and (Ba, Rb)122 (Fig. 10). In (Ba, Rb)122 the presence of magnetic fields shift the onset of superconductivity to lower temperatures, but do not cause any broadening. This shift of $T_c$ is linear in the applied magnetic field and therefore the upper critical fields $H_{c2}^{||ab}$ and $H_{c2}^{||c}$ do not show any significant curvature. Sm1111, however, shows a distinctly different behavior. Magnetic fields cause only a slight shift of the onset of superconductivity, but a significant broadening of the transition, indicating weaker pinning and accordingly larger flux flow dissipation. The resistivity in our Sm1111 crystals shows typically two or three steps at low fields. These steps vanish at magnetic fields higher than 1 T.

The diamagnetic signal does not show any steps, thus they may be associated with surface imperfections. While (Ba, Rb)122 shows a sharp transition and a clear onset of superconductivity, there is no sharp transition in Sm1111 and $T_c$ is therefore less clearly defined. We chose three different criteria: 90% $\rho_n$, 50% $\rho_n$ and 10% $\rho_n$, where $\rho_n(T)$ is the linear extrapolation of the normal state resistivity. Due to the absence of field-induced broadening in (Ba, Rb)122, all definitions of $T_c$ lead to the same results for the $H_{c2}$ slopes and we chose the 50% $\rho_n$ criterion. The upper critical fields $H_{c2}^{\|ab}$ and $H_{c2}^{\|c}$ extracted from resistivity measurements of both materials are shown in Fig. 11. The upper critical fields $H_{c2}$ in (Ba, Rb)122 increase linearly ~0.5K below Tc, with a slope of 4.2 T/K ($H\|c$) and 7.1 T/K ($H\|ab$). The slopes $dH_{c2}/dT$ in Sm1111 depend strongly on the choice of the criterion of superconductivity due to the pronounced broadening. We found 3.3 – 1.2 T/K for $H\|c$ and 8.0 – 5.5 T/K for $H\|ab$. These large slopes already indicate very high values of $H_{c2}(0)$.

The (Ba, Rb)122 structure is more isotropic than the structure of Sm1111, which is already manifested in the upper critical field anisotropy $\gamma_H = H_{c2}^{\|ab} / H_{c2}^{\|c}$ (insets Fig. 11). The anisotropy $\gamma_H$ in Sm1111 ranges between 7 – 7.5 using the 50% $\rho_n$ criterion, while it was found to be between 2.5 - 3.2 in (Ba, Rb)122. This reflects a stronger coupling of the FeAs-layers and therefore more electronic coupling in the A122 compounds compared to the Ln1111 compounds. The anisotropy $\gamma_H$ is temperature dependent and decreases with decreasing temperature for both (Ba, Rb)122 and Sm1111. This is in very good agreement with previous experiments and seems to be a general feature in all different classes of pnictide superconductors.

**ii. Magnetic measurements and critical current density**

Temperature dependence of the magnetic moment, measured in a magnetic field of 1 mT parallel to the *c*-axis for a single crystal of SmFeAsO$_{0.6}$F$_{0.35}$ with a mass of about 6 μg, is presented in Fig. 12. The sharp transition to the superconducting state is characteristic for a high quality single crystal. A transition temperature of 52 K indicates that the crystal is close to optimal doping. The value of the zero field cooled magnetic moment reflects the full diamagnetic response of the crystal studied. The small ratio of field cooled to zero field cooled magnetization is characteristic for a superconductor with relatively strong pinning, which was confirmed in the magnetic hysteresis loop measurements reported previously [50]. A wide loop measured at 5 K in a magnetic field up to 7 T revealed a hysteresis width almost independent on the field. A high critical current density of the crystal was deduced, reaching values of about $10^{10}$ A/m$^2$ at 5 K. It was suggested that a slight increase of the critical current density for higher magnetic field may indicate the increase of the effectiveness of pinning centers with increasing magnetic field [45]. The critical current density at 2 K, 5 K and 15 K, estimated from the field dependence of the magnetic moment for SmFeAsO$_{0.8}$F$_{0.2}$, are higher than $10^9$ A/m$^2$ (Fig. 13), what is promising for having applications in mind. Very similar results were obtained for single crystals of Ba$_{0.89}$Rb$_{0.05}$Sn$_{0.06}$ Fe$_2$As$_2$. The observed extremely small Meissner fraction is due to the pronounced magnetic irreversibility, stemming most likely from the lattice mismatch caused by the substitution with relatively big Rb ions, introducing effective pinning centers. Again, a relatively strong pinning was confirmed in magnetic hysteresis loop measurements [55]. The critical current density at 2 K and 5 K reaches values of the order of $10^{10}$ A/m$^2$, similar like the Sm1111 results. The slight increase of the critical current density with increasing field was observed and it is most likely due to the occurrence of the peak effect.

### iii. Magnetic torque investigations

First, the analysis of the measured torque data was done with the simple one-anisotropy model. The anisotropy parameter was found to be up to 1.4 T almost field independent, but varies strongly in temperature between 8 at $T \approx T_c$ and 23 at $T \approx 0.4T_c$ for a SmFeAsO$_{0.8}$F$_{0.2}$ crystal with $T_c$ of 45 K [31]. This disagrees with the temperature dependence of $\gamma_H$ determined from $H_{c2}$ by resistivity measurements as shown above, and with recent values of $\gamma_H$ of NdFeAs(O, F) crystals obtained from high field resistivity measurements [33]. It is evident that $\gamma_H$ decreases with decreasing temperature and with temperature dependent values $\gamma_H$ all much smaller than the torque results.

Nevertheless, a temperature dependent $\gamma$ would imply an unconventional (non – Ginzburg–Landau) behavior of the thermodynamic parameters. A possible and natural explanation would be multi–band superconductivity, where different parts of the Fermi surface sheet develop distinct gaps in the superconducting state. The complex interband and intraband scattering of charge carriers will rule the physics and therefore influence strongly the superconducting state anisotropy. In this framework a temperature and even field dependent anisotropy can be well understood, since interband and intraband scattering processes will lead to modifications in the simple one gap Ginzburg–Landau relation.

Since for multi band superconductors the magnetic penetration depth anisotropy $\gamma_\lambda$ might be different from the upper critical field anisotropy $\gamma_H$ a more general two-anisotropy model for the angular dependent torque was proposed by Kogan [26, 61]. We performed a detailed analysis of several SmFeAsO$_{0.8}$F$_{0.2}$ and NdFeAsO$_{0.8}$F$_{0.2}$ single crystals by torque experiments using the two-anisotropy model. Details of the calculations have been published elsewhere [62]. By fixing the upper critical field anisotropy $\gamma_H$ to the values obtained by Jaroszynski *et al*. [33] the magnetic penetration depth anisotropy $\gamma_\lambda$ was found to be strongly temperature dependent in a very similar way as the single anisotropy model predicted. Figure 14 shows exemplarity one sketch of angular dependent magnetic torque experiment. Figure 15 shows the extracted $\gamma_\lambda$ and $\gamma_H$ as a function of temperature, derived by systematic fitting of the torque data.

As a result, the low field torque is mostly sensitive to the magnetic penetration depth anisotropy $\gamma_\lambda$ and is almost insensitive on the upper critical field anisotropy $\gamma_H$. This might be different in higher fields close to the phase transition, where the effective anisotropy of the system would correspond much more to the upper critical field anisotropy, which should lead to a reduction of $\gamma$ in higher fields [32]. For the temperature dependence of the magnetic penetration depth a good agreement was found with the two fluid model [63] and the value of $\lambda_{ab}(0) = 250$ nm extracted from the data is in good agreement with other estimates [64-66]. The upper critical field was estimated according to the WHH relation [67] to be $\mu_0 H_{c2} = 70$ T.

We would like to stress, that crystals of a bad quality, containing domains with different crystallographic orientation show a much lower and temperature independent anisotropy. This might be a reason for the various results on the magnetic penetration depth anisotropy published recently by several groups. To illustrate this, Fig. 16 presents torque measurements performed on a crystal which shows a broad superconducting transition and slight disorder in the structure observed by X-ray diffraction. The anisotropy is strongly reduced. Therefore, well characterized single crystals of high quality are better suited to study anisotropic properties.

## 3. Point-contact Andreev-reflection spectroscopy measurements

Figure 17 a)-c) shows some examples of low-temperature normalized conductance curves (symbols) measured on various Au/SmFeAsO$_{1-x}$F$_y$ junctions. The enhancement of the conductance around zero bias and the presence of peaks clearly reveal the occurrence of the Andreev-reflection phenomenon at the N/S junction. The peaks in the conductance curves indicate the presence of a superconducting gap while higher-bias features (such as a widening of the Andreev feature) in Fig. 17 a) and b) and small humps (indicated by arrows) in the curve shown in Fig. 17 c) suggest the presence of a second gap. In order to confirm this observation a comparison has been carried out between the one-gap and two-gap models for Andreev reflection at the N/S interface. Since no zero bias conductance peaks (ZBCP) are present, the fitting has been performed using nodeless gaps. The modified [68] Blonder-Tinkham-Klapwijk (BTK) model [69] generalized to take into account the angular distribution of the current injection at the interface [70] was adopted. In the two-band case the conductance is the weighed sum of two BTK contributions: $G = w_1 G_1 + (1 - w_1) G_2$ [67].

The result of the comparison is also shown in Fig. 17 a)-c): the one-band model (blue dash lines) reproduces only a small central portion of the curves while the two-band model (red solid lines) is remarkably better, indicating the presence of two superconducting gaps whose values, as determined by the fitting procedure, are $\Delta_1 = 6.45 \pm 0.25$ meV and $\Delta_2 = 16.6 \pm 1.6$ meV, with ratios $2\Delta_1/k_B T_c$ = 2.8–3.1 and $2\Delta_2/k_B T_c$ = 6.8–8.5, in good agreement with PCAR spectroscopy results on polycrystalline samples [47]. Figure 18 a) shows the temperature dependence of the raw conductance curves of the same Au/SmFeAsO$_{0.8}$F$_{0.2}$ junction whose low-temperature curve is shown in Fig. 17 c). The overall appearance of the conductance curves is asymmetric, being higher at negative bias voltages, as it has also been observed in several other PCAR spectroscopy measurements in iron pnictides. Furthermore, the Andreev peaks in the low-temperature conductance usually are of opposite asymmetry (i.e. they are higher at positive bias voltage). This asymmetry is more or less pronounced. When this feature is significant, a separate fit of negative and of the positive bias part has been performed (Fig. 17 c) and 18 b)), but further studies have to be carried out in order to clarify its origin.

As a further confirmation of the presence of two gaps in SmFeAsO$_{1-x}$F$_y$, a fit of the temperature dependence of the conductance curves has been carried out. Figure 18 b) shows the temperature dependence of the normalized conductance curves (symbols) shown in panel a) together with their relevant two-band BTK fitting curves (lines). Since the asymmetry is rather pronounced in this case, the left and the right part of the curves are fitted with slightly different parameters. The gap values obtained by this procedure are reported in the Fig. 18 c). As far as the small gap, $\Delta_1$ is concerned, the values are almost identical in the two cases (open and full circles, respectively) while a small difference is derived for the larger one (open and full squares, respectively). Both $\Delta_1$ and $\Delta_2$ follow a BCS-like behavior (blue dash and red dash-dot lines, respectively) and close at the critical temperature of the junction.

**4. Scanning Tunneling Spectroscopy measurements**

The surfaces of SmFeAsO$_{0.8}$F$_{0.2}$ crystals imaged by scanning tunneling microscopy [71] present plateaus of irregular shape, with an rms roughness of less than 0.1 nm. These terraces have a height difference of 0.8 or 1.6 nm steps, roughly a single or double crystal *c*-axis lattice constant for SmFeAsO$_{0.8}$F$_{0.2}$ (*c*=0.847 nm).

Low temperature tunneling conductance spectra are shown in Fig. 19. These data (a and b) have been acquired at two different locations of the sample surface, using the same tunneling conditions. At energies close to the Fermi level (low bias voltage), the d*I*/d*V* spectra present a conductance depletion. At the edge of this depletion, conductance kinks or faint peaks are detected, indicated in the figure by the dotted lines. Following common practice [72], we consider half the peak-to-peak energy separation as a measure of the superconducting gap. Considering thousands of spectra acquired within regions of tens of nanometer wide, we find a mean gap value of 7(1) meV. The variation in Δ (~14%) is relatively small compared to the gap distributions usually measured in Bi-based high-$T_c$ cuprates [72]. The average critical temperature of several samples of the same batch is $T_c$ = 45 K, yielding 2Δ/k$_B T_c$=3.6, a ratio close to that expected for a weak-coupling s-wave superconductivity.

A second gap-like feature is detected at voltages around 20 meV (see arrows in Fig. 19 a and b). In contrast to the low energy feature, the peaks at higher energy vary in height, are much wider, and are located over a broader energy scale. Remarkably, they are often not detected simultaneously for occupied and empty states, and when they are, their energy locations are not symmetric with respect to the Fermi level. The last fact questions the interpretation of the high-voltage feature as a second superconducting gap.

A tunneling conductance spectrum over a wider voltage range is shown in Fig. 19c. The conductance at high bias voltages is voltage dependent (*V*-shaped) with a strong particle-hole asymmetry, the conductance measured at negative sample bias (occupied states) being systematically higher than the conductance at positive sample bias (empty states). This asymmetry is strikingly similar to the one measured in a number of high-$T_c$ cuprates [72], possibly indicating strong electronic correlations in this compound.

The 7 meV value found for the low energy gap is in agreement with the values measured in point contact spectroscopy studies on similar compounds and on polycrystalline samples [47, 73]. Moreover, the value of the 2Δ/k$_B T_c$ ratio suggests that this spectroscopic feature is the signature of a superconducting gap. However, caution imposes on the interpretation of the high-energy feature observed in SmFeAs(O, F) as a second superconducting gap. Although the energy scale of this feature is of the order of the one reported in point contact spectroscopy [47, 73], it is not systematically detected for empty and occupied states and it is particle-hole asymmetric. Moreover, it is not systematically detected in a recent STS study of $BaFe_{1.8}Co_{0.2}As_2$ single crystals [74]. Therefore the STS measurements rather cast doubts on this feature being a manifestation of a second superconducting gap. In order to elucidate this discrepancy on the interpretation of the data, detailed studies of the temperature dependence of the spectral features, vortex core spectroscopy and/or tunneling along different crystallographic directions are needed.

**IV. Conclusions**

Single crystals of $LnFeAsO_{1-x}F_x$ (Ln=La, Pr, Nd, Sm, Gd) have been grown using a cubic anvil high pressure technique, and $Ba_{1-x}Rb_xFe_2As_2$ crystals have been grown in quartz ampoules. Superconductivity in the Ba122 compound has been induced by Rb substitution for the first time. The availability of Ln1111 single crystals made it possible to determine several basic superconducting parameters, such as upper critical fields and their anisotropy $\gamma_H$ and magnetic penetration depth anisotropy $\gamma_\lambda$. The anisotropy $\gamma_\lambda$ is temperature dependent and increases with decreasing temperature from $\gamma_\lambda(T_c) = \gamma_H(T_c)=7$ towards $\gamma_\lambda(0)=19$, while the anisotropy $\gamma_H$ varies much less and decreases towards $\gamma_H(0)=2$ with decreasing temperature [62]. This is suggestive of two superconducting gaps, similarly to the situation in $MgB_2$. PCAR studies support this scenario and show the existence of two gaps, $\Delta_1 = 6.45 \pm 0.25$ meV and $\Delta_2 = 16.6 \pm 1.6$ meV, in good agreement with results on polycrystalline samples [47]. STM investigations reveal a superconducting gap at the energy scale of 7(1) meV and a high-energy feature around 20 meV whose connection to a second superconducting gap has to be further explored. The critical current is relatively high, with $J_c$ values of $2\times10^9$ A/m$^2$ at 15K in field up to 7 T. $Ba_{1-x}Rb_xFe_2As_2$ crystals are electronically more isotropic, indicative of better coupling of the FeAs layers by the (Ba, Rb) layers than by the Sm(O, F) layers

*Acknowledgements*

This work w as supported by the Swiss National Science Foundation, by the NCCR program MaNEP, and partially supported by the Polish Ministry of Science and Higher Education under research project for the years 2007-2009 (No. N N202 4132 33).

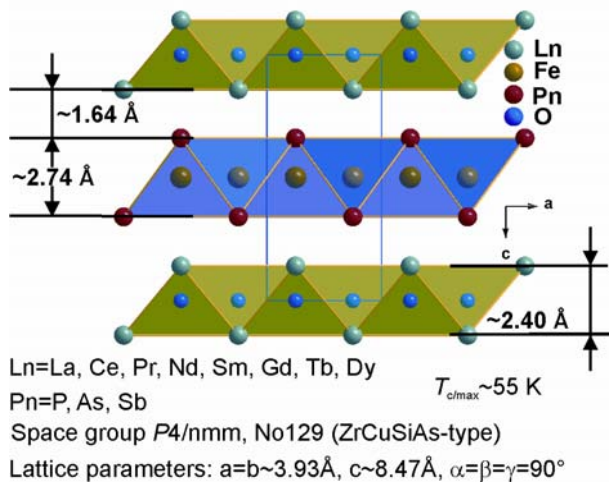

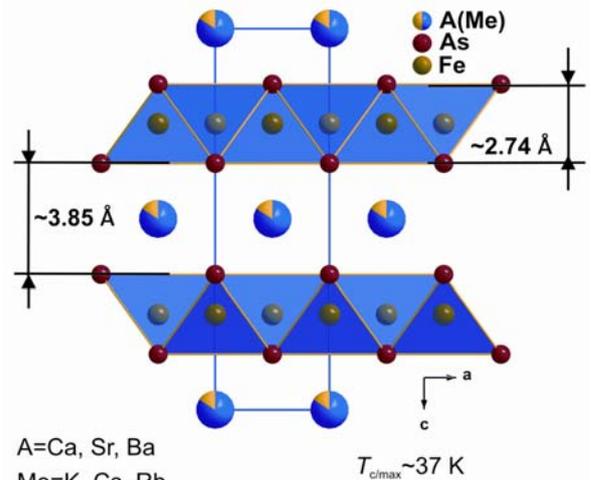

Fig. 1. Structure projection of Ln1111 along the *b* direction.

Fig. 2. Structure projection of A122 along the *b* direction.

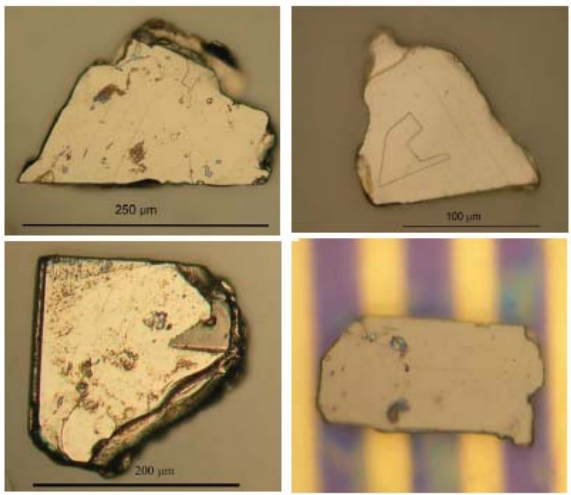

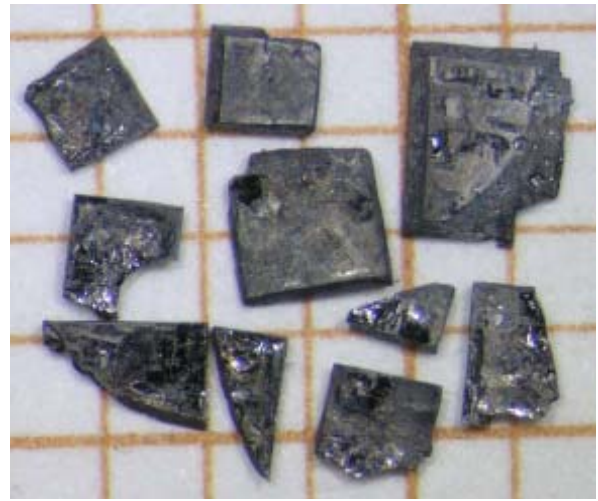

Fig. 3. Single crystals of $SmFeAsO_{1-x}F_y$.

Fig 4. Single crystals of $Ba_{0.9}Rb_{0.1}Fe_2As_2$ on a millimeter grid.

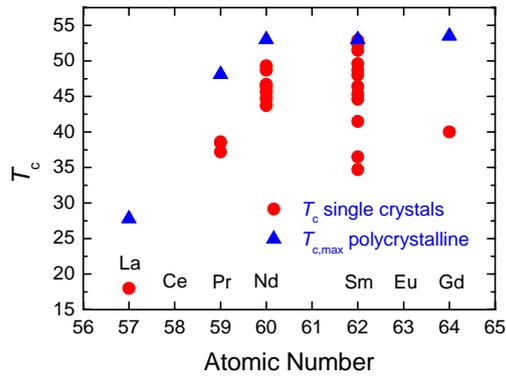
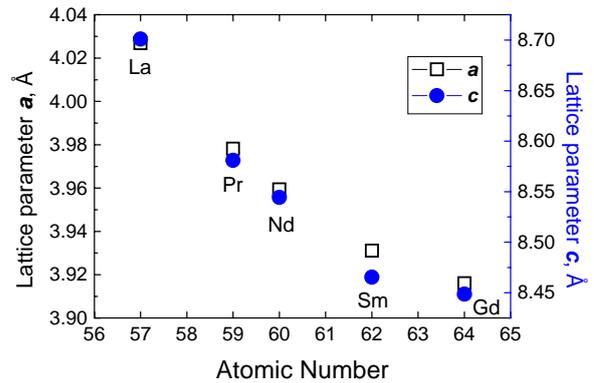

Fig. 5. $T_c$ measured on single crystals from various growth experiments with different doping level, compared with $T_{c,max}$ from literature measured on polycrystalline samples vs. the atomic number.

Fig. 6. Cell parameters as a function of the atomic number for LnFeAs(O,F) (Ln=La, Pr, Nd, Sm, Gd). Squares show the *a* and circles represent the *c* lattice parameters.

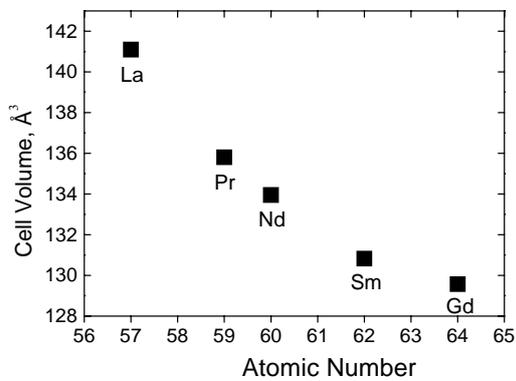
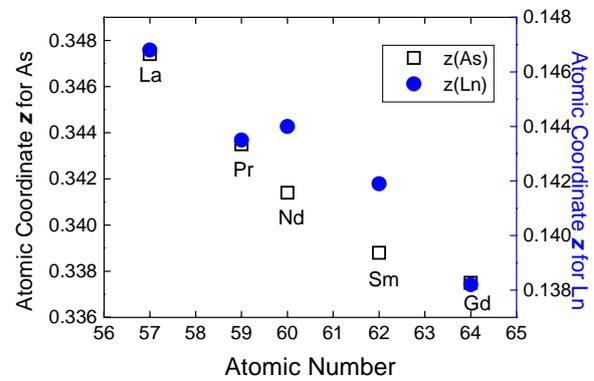

Fig. 7. Cell volume as a function of the atomic number for LnFeAs(O,F) (Ln=La, Pr, Nd, Sm, Gd).

Fig. 8. Atomic coordinates *z* as a function of atomic number. Squares show *z* for As and circles represent *z* for LnFeAs(O,F) (Ln=La, Pr, Nd, Sm, Gd).

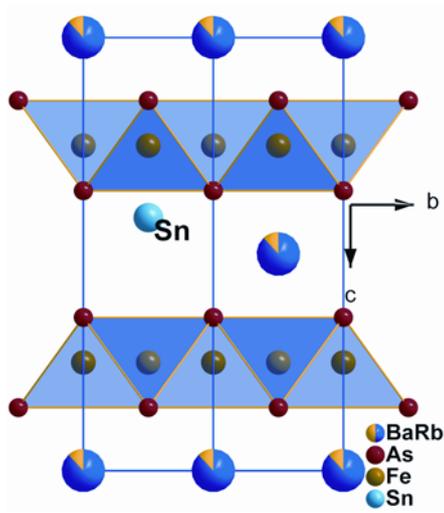

Fig. 9. Crystal structure of $Ba_{0.89}Rb_{0.05}Sn_{0.06}Fe_2As_2$.

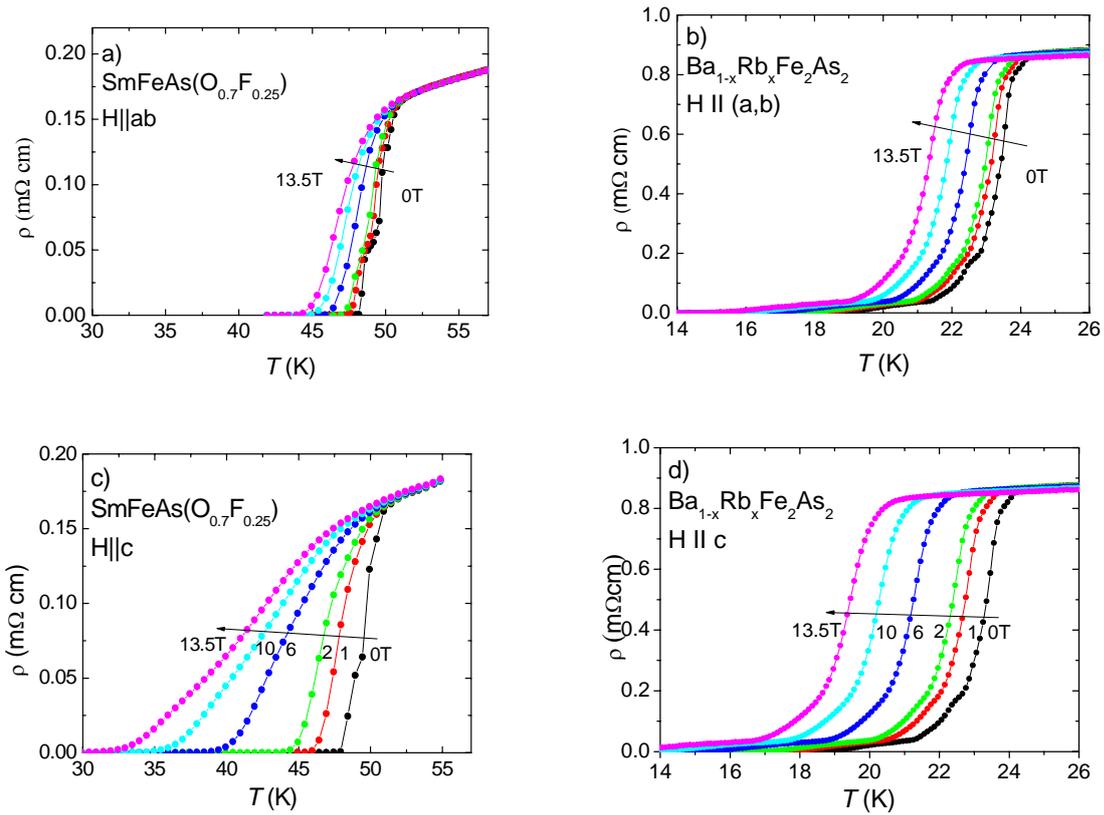

Fig 10. Examples of resistivity $\rho(T,H)$ for $SmFeAsO_{0.8}F_{0.2}$ and for $Ba_{0.89}Rb_{0.05}Sn_{0.06}Fe_2As_2$ single crystals measured in fields applied parallel to the ($Fe_2As_2$)-layers ($H\|ab$) (a and b) and perpendicular to them ($H\|c$) (c and d), at magnetic field strengths of 0, 1, 2, 6, 10 and 13.5 T.

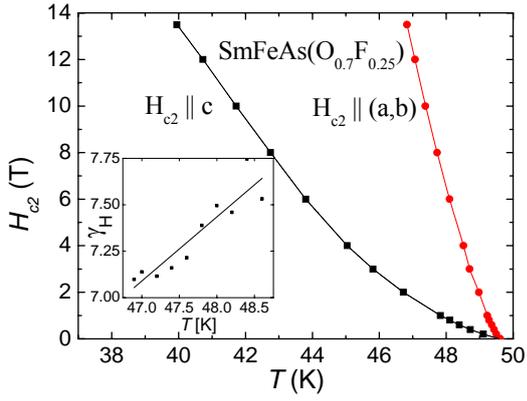 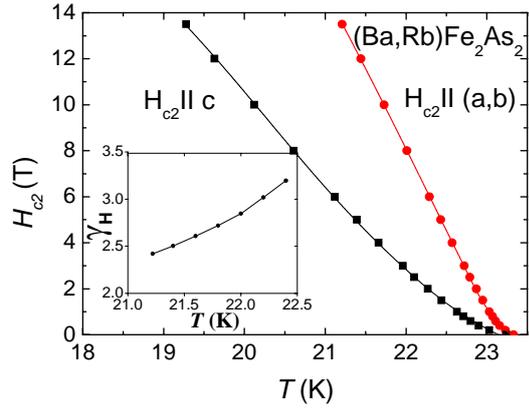

a) b)

Fig 11. Temperature dependence of the upper critical field with $H_{c2}^{\parallel ab}$ and $H_{c2}^{\parallel c}$ for the SmFeAs(O,F) and the $(Ba,Rb)Fe_2As_2$ system. Inset: The upper critical field anisotropy $\gamma_H = H_{c2}^{\parallel ab}/H_{c2}^{\parallel c}$ in the vicinity of $T_c$.

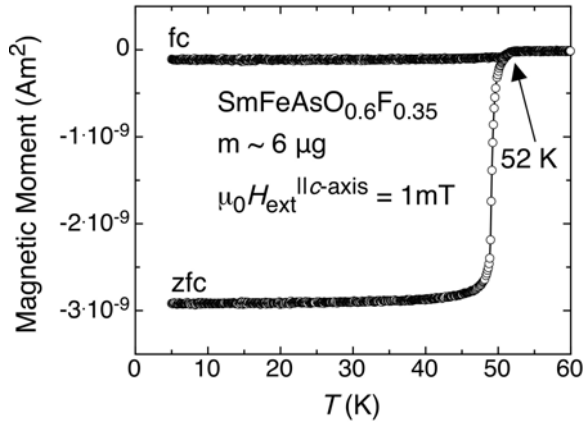 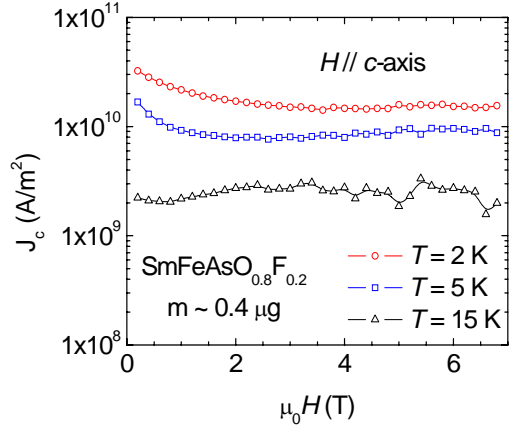

Fig. 12. Temperature dependence of the magnetic moment measured on a $SmFeAsO_{0.6}F_{0.35}$ (nominal content) single crystal with $T_c$ of 52 K in an applied field of 1 mT parallel to its *c*-axis. ZFC and FC denote zero-field cooling and field cooling curves, respectively.

Fig. 13. Critical current density calculated from the width of the hysteresis loop measurements up to 7 T at 2 K, 5 K and 15 K.

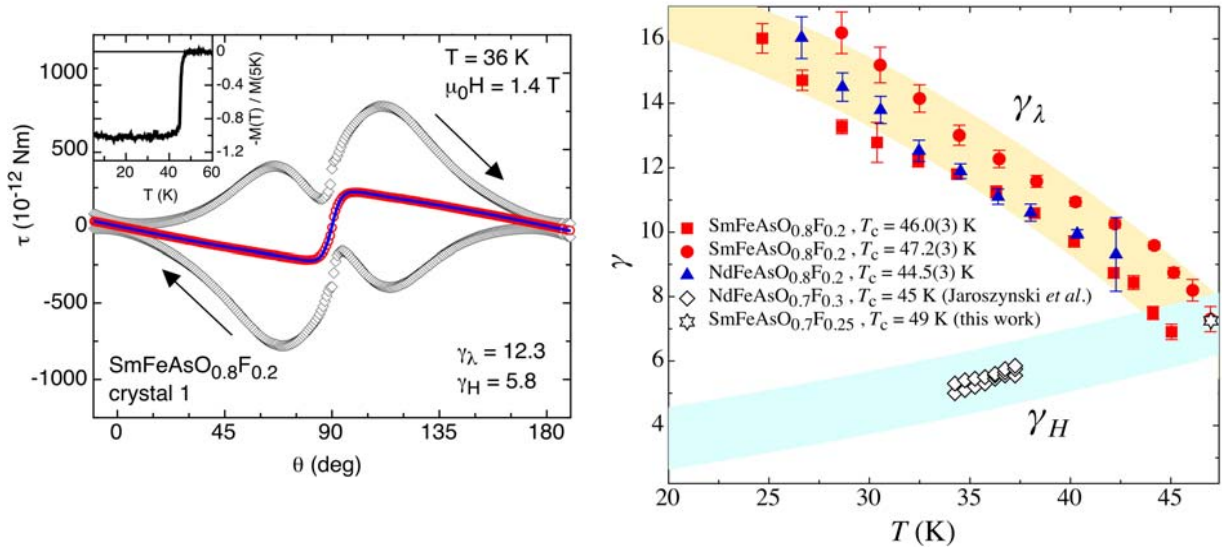

Fig. 14. Angle dependent raw torque measured for a SmFeAsO$_{0.8}$F$_{0.2}$ single crystal at 36 K in 1.4 T. The squares denote the raw (irreversible) torque, which was subsequently averaged in order to obtain the reversible component (red circles). The blue line is a fit with the parameters $\gamma_\lambda$ = 12.3 and $\gamma_H$ = 5.8. The inset displays the low field magnetic moment with a sharp superconducting transition, suggestive for the excellent quality of the single crystal.

Fig. 15. Summary of the anisotropies $\gamma_\lambda$ deduced from the torque data, using fixed values for $\gamma_H$ after Jaroszynski *et al*. [33]. ☆ - $\gamma_H$ determined from Fig.11b). All data plotted in this figure are described in more detail in [62].

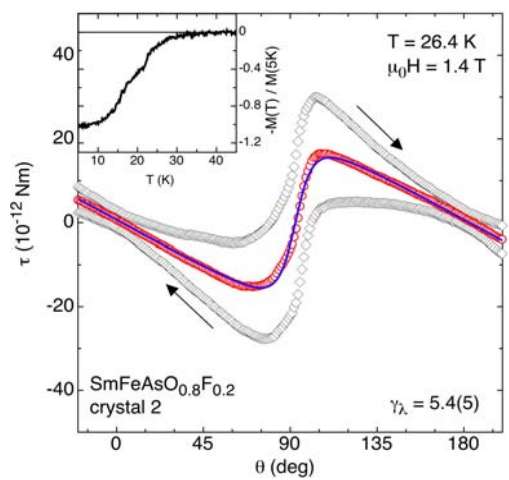

Fig. 16. Torque data derived on a SmFeAsO$_{0.8}$F$_{0.2}$ crystal of inferior quality. The usually sharply featured angular torque is distorted in the *ab*-plane (90 degree), just where the fitting curve is mostly sensitive to the anisotropy parameter $\gamma_\lambda$. The fitted $\gamma_\lambda$ is found to be strongly reduced 5.4(5). The inset shows the broad transition of the low field magnetic moment, measured in a SQUID experiment.

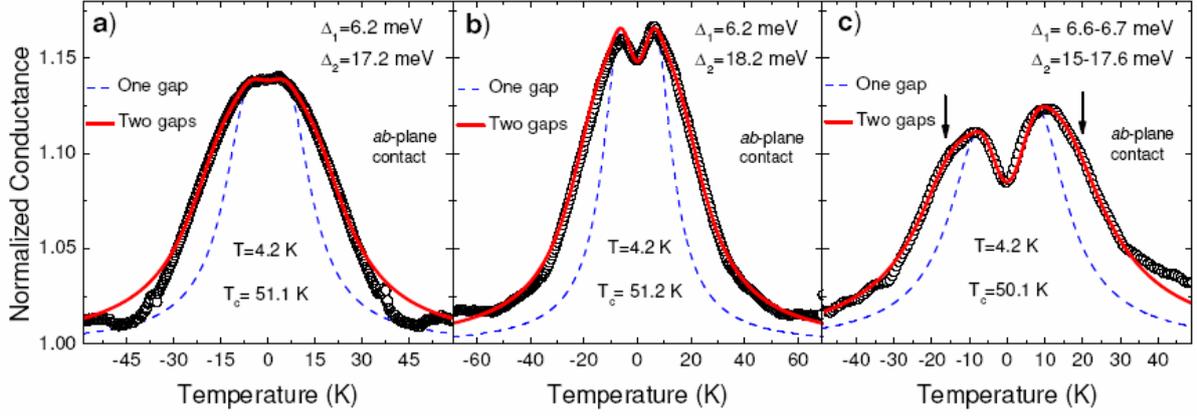

Figure 17, a)-c): examples of normalized conductance curves measured at 4.2 K in Au/SmFeAsO$_{0.8}$F$_{0.2}$ junctions (symbols). The main direction of current injection is along the *ab*-planes of the crystals (*ab*-plane contacts). Blue dash lines: single-band BTK fit. Red solid lines: two-band BTK fit.

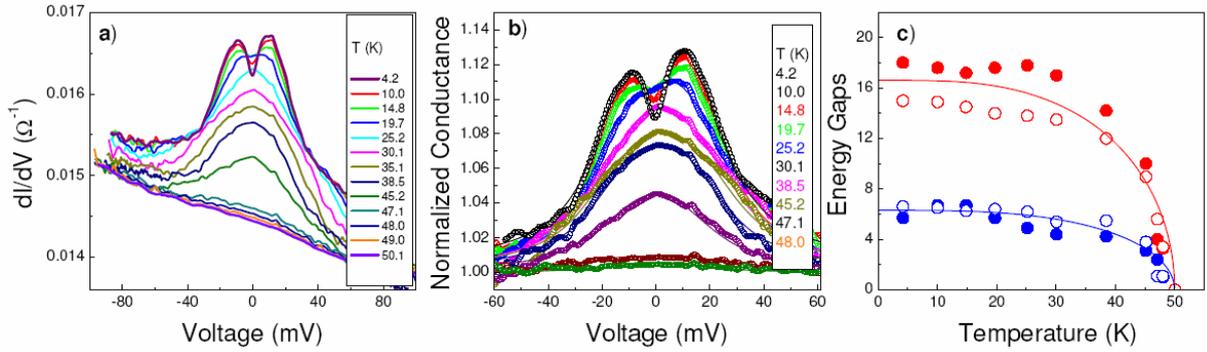

Figure 18. a): Temperature dependence of the raw conductance curves for a Au/SmFeAsO$_{0.8}$F$_{0.2}$ single crystal point-contact junction with the current mainly injected along the *ab*-planes. b): temperature dependence of the normalized conductance curves shown in a) (symbols) together with their relevant two-gap BTK fits (lines). c) gap values obtained by fitting the curves shown in b). Due to the asymmetry of the conductance curves, the extracted gap values differ slightly between the negative (open symbols) and the positive bias part (full symbols).

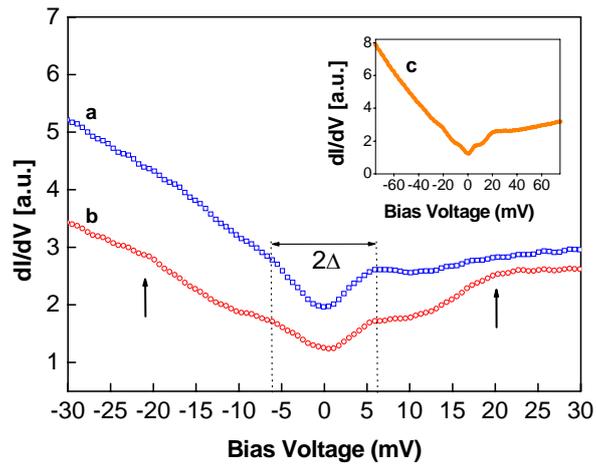

Figure 19. a and b : Scanning tunneling spectra measured on a SmFeAsO$_{0.8}$F$_{0.2}$ single crystal at 4.2 K, at two different locations of the sample. The dotted lines are set at the position of the peaks or kinks used for determining the value of the superconducting gap 2Δ. The arrows indicate the position of the gap-like features detected at higher bias. c : Local tunneling spectrum displayed over a wide energy range: the voltage-dependent conductance at high bias and the particle-hole asymmetry can be interpreted as indications of strong electronic correlations.